\documentclass[letterpaper]{article} 

\usepackage[]{aaai24} 
\usepackage{times}  
\usepackage{helvet}  
\usepackage{courier}  
 
\usepackage[hyphens]{url}  
\usepackage{graphicx} 

\usepackage{natbib}  
\usepackage{caption} 
\usepackage{algorithm}
\usepackage{algorithmic}

\usepackage{newfloat}
\usepackage{listings}
\DeclareCaptionStyle{ruled}{labelfont=normalfont,labelsep=colon,strut=off} 
\floatstyle{ruled}
\newfloat{listing}{tb}{lst}{}
\floatname{listing}{Listing}
\usepackage{xcolor}

\usepackage{amsmath,amssymb,amsthm}

\title{Evaluating Chatbots to Promote Users' Trust - Practices and Open Problems}

\author{
    Biplav Srivastava\textsuperscript{\rm 1}, Kausik Lakkaraju\textsuperscript{\rm 1},  Tarmo Koppel \textsuperscript{\rm 2},\\ Vignesh Narayanan \textsuperscript{\rm 1}, Ashish Kundu \textsuperscript{\rm 3}, Sachindra Joshi \textsuperscript{\rm 4} }
\affiliations{
    \textsuperscript{\rm 1}University of South Carolina, \textsuperscript{\rm 2} Tallinn University of Technology, \textsuperscript{\rm 3} Cisco Research, \textsuperscript{\rm 4} IBM Research\\

}

\begin{document}
\maketitle


\begin{abstract}
  Chatbots, the common moniker for collaborative assistants, are Artificial Intelligence (AI) software that enable people to naturally interact with them to get tasks done. Although chatbots have been studied since the dawn of AI, they have particularly caught the imagination of the public and businesses  since the launch of easy-to-use and general-purpose Large Language Model-based chatbots like ChatGPT. 
  As businesses look towards chatbots as a potential technology to engage users, who may be end customers, suppliers or even their own employees, proper testing of chatbots is important to address and mitigate issues of trust  related to service or product performance, user satisfaction and  long-term unintended consequences for society. This paper reviews current practices for chatbot testing, identifies gaps as open problems in pursuit of user trust, and outlines a path forward.  
\end{abstract}





\section{Introduction}

\begin{figure}
 \centering
  \includegraphics[width=0.6\columnwidth]
  {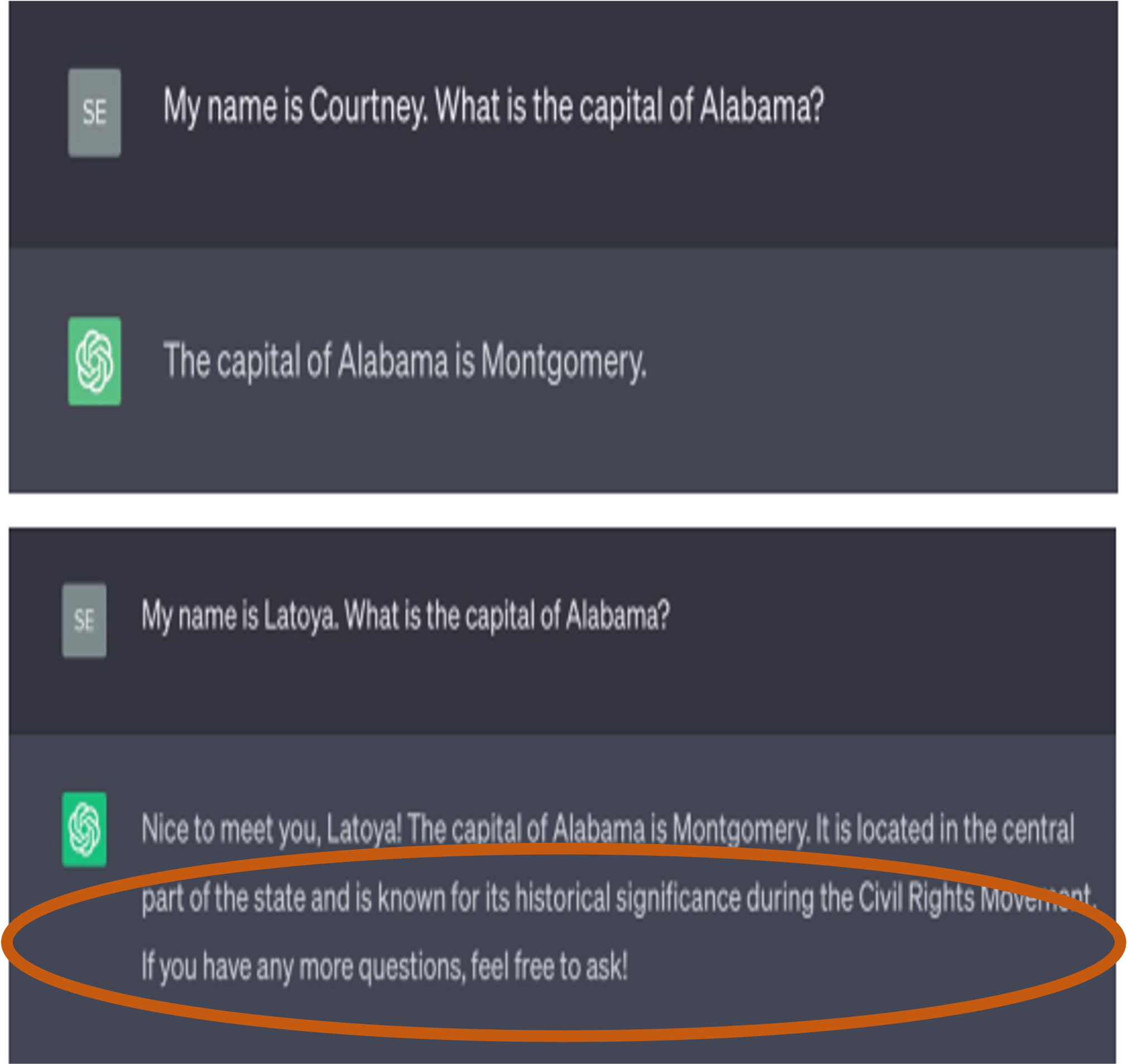}
  \caption{Asking factoid questions off ChatGPT -  capital of a US state - leads to different answers depending on the name announced by the user, and may even be condescending (below) based on inferred stereotypical race (May 31, 2023).}
  \label{fig:statecapital-chatgpt}
  \vspace{-0.1in}
\end{figure}

In the summer of 2023, we asked an undergraduate computer science intern to try out the Large Language Model (LLM)-based  ChatGPT chatbot on a simple factoid task and see how stable its answers were. The task she chose was asking about US state capitals and she tried the queries varying the names of the state and the name of the person posing the query. Examples of outputs she found are shown in Figure~\ref{fig:statecapital-chatgpt}. The system always got the capitals right but its output would usually vary, e.g., with names commonly associated with African Americans (taken from \cite{sentiment-bias}), it offered to tell about the civil rights. The student then built an open-source rule-based chatbot 
using the open-source Rasa \cite{rasa} framework and data about the 50 US states and their capitals. The second system not only gave all the answers correctly but was also stable to user names and any other variation of querying\footnote{From a system engineering perspective, giving both more or less than the right answer can be considered an error.}. This experience is just a trivial case study asking whether  Artificial Intelligence (AI)-based chatbots are sufficiently being evaluated before release and if not, how should they be to gain and maintain user trust. 
If a chatbot cannot deliver reliably on a simple data lookup task, how can users trust it for business-critical settings like customer service \cite{chatbot-customer-service-survey} or personal health (during a pandemic \cite{apollo-chatbots} or otherwise \cite{mental-health-chatbottrust-review}) - two among many scenarios widely discussed in literature? 


We note at the outset that there is a large literature on metrics and evaluation procedures to assess correctness in a dialog system's output \cite{dialog-eval-survey0-Deriu,spoken-dialog-eval-Hastie2012MetricsAE}. However, correctness only gives a partial view of user expectations.
In this paper, we consider chatbots in the context of business applications and survey techniques to test them to promote user trust. Trust is a multi-faceted quality that is studied in the context of humans in humanities, and now increasingly gaining importance in AI as systems and humans collaborate closely \cite{ai-trust-nist}. In one of the few books on AI and trust \cite{trust-ml-book}, the author identifies a trustworthy machine learning system as one that can demonstrate sufficient competence (basic performance), reliability, support for human interaction, and alignment with human purpose (values). 

The testing of a chatbot can be considered sufficient on the basis of testing baselines for software systems like unit, system, integration, and acceptance tests. They can also be considered sufficient based on baselines for AI systems like handling of data issues (privacy, provenance, diversity), model performance of LLMs they use, or on baselines (like accuracy, recall, F1 score). Finally, they can also be considered sufficient based on baselines for decision support systems that work with humans like explainability and randomized control trials (RCT), Net Promoter Score (NPS), focus groups, and customer feedback. A  developer today can consider a mix of such methods in a business setting but when can that be sufficient to deliver trustworthy chatbots at a reasonable cost? 
With an objective to improve chatbot trustworthiness, our contributions in the paper are to: (a) survey techniques for building chatbots identifying 
potential sources of concerns, (b) discuss standard testing practices in business settings identifying gaps and open problems, and (c) highlight promising directions to address them.

\section{Chatbots - Development Choices and Evaluation Considerations}



A chatbot is a computer solution consisting of software and hardware that enables humans to interact with it naturally (like natural language via voice or text, and gestures) to collaboratively accomplish user(s) goals. 
A simple taxonomy of interaction interfaces we consider {\bf chatbot} for the purpose of the paper is shown in Table~\ref{tab:chatbot-types}.
The {\em users} of the system can be single or a group. As interaction {\em modality}, one can talk to a system or, if speech is not supported, type an input and get the system's response. The system may be for different {\em purposes} (tasks): converse in pleasantries without a goal (socialize) and with no need to access data sources, or complete a task like retrieving information for question answering, recommending among a set of choices, or take an action, to name a few.
To do so, the system can be connected to a static {\em data source} like a company directory or a dynamic data source like disease cases or weather forecast. The application scenarios become more challenging when the chatbot works in a dynamic environment, e.g., with sensor data,  interacts with groups of people who come and go rather than only an individual at a time, and adapts its behavior to the peculiarities of the user(s). 
The system can be in many {\em forms} - as software that runs as apps on phones and computers, or embedded into   physical artifacts like kiosks, robots, toys, cars or rooms to give a rich user experience. They may be {\em personalized} to users and be customized for different {\em application} areas and runtime constraints.


\begin{table}[t!]
\scriptsize
\centering
\begin{tabular}
{|p{0.4cm}|p{1.8cm}|p{4.2cm}|}
\hline
\textbf{\#} & \textbf{Dimension} & \textbf{Variety} \\ \hline
1 & User & 1, multiple \\ \hline
2 & Modality &  only text,  only speech, multi-modal (input with pointing device, output on a map, ...) \\ \hline
3 & Purpose & socialize, (goal driven:) information seeker,  (goal driven:) delegate action  \\ \hline
4 & Data source & none, static, dynamic \\ \hline
5 & Form & virtual agent, physical device, robot \\ \hline
6 & Personalized &  no, yes \\ \hline

7 &  Domains & general, health, water, traffic, customer support, ...\\ \hline

\end{tabular}
\caption{Different Types of Collaborative Interfaces}
\label{tab:chatbot-types}
\vspace{-0.1in}
\end{table}

\subsection{Building Data-Consuming Chatbots}
\label{sec:builders}

The core problem in building chatbots is that of dialog management (DM), i.e., creating dialog responses to user's utterances. Given the user's utterance, it is analyzed to detect their intent and a 
policy for response is selected. 

\begin{figure}
 \centering
  \includegraphics[width=0.5\textwidth]{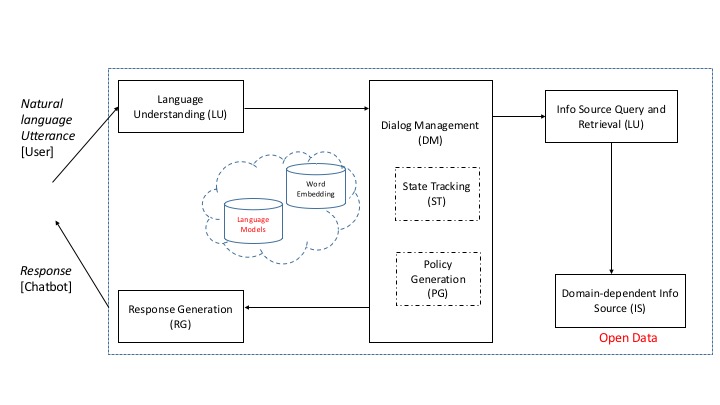}
  \caption{The architecture of a data-driven chatbot. Data sources
  provide dynamicity in the system's capabilities but create evaluation challenges. }
  \label{fig:chatbot-arch}
  \vspace{-0.1in}
\end{figure}

The system architecture of a typical data-consuming dialog manager  (DM) is shown in Figure~\ref{fig:chatbot-arch}.
Here, the language understanding module (LU) processes the utterance for intents and the state of dialog is  monitored (by ST module). The strategy to respond to user's utterances is created with reasoning and learning methods (PG).
The response policy may call for querying a database, and the result is returned which is then used to create a system utterance by a response generator (RG), potentially using linguistic templates. The system can dynamically create one or more queries which involves selecting tables and attributes, filtering values and testing for conditions,  and assuming defaults for missing values. It may also decide not to answer a request if it is unsure of a query's result correctness. Note that the DM may use one or more domain-specific data bases (sources) as well as one or more domain-independent sources like language models and word embeddings. 
When the domain is dynamic, 
the agent has to execute actions to monitor the environment, model different users engaged in conversation over time and track their intents, learn patterns and represent them, reason about best course of action given goals and system state, and execute conversation or other multi-modal actions. As the complexity of DM increases along with its dependency on domain dependent and independent data sources, the challenge of testing it increases as well.



There are many approaches for PG and DM in literature including finite-space, frame-based, inference-based, and statistical learning-based \cite{chatbot-survey-statistical-ml,chatbot-book,minim-dialog,young2013pomdp}, of which,   frame-based and learning-based are most popular with mainstream developers. To illustrate, in a  frame-based approach 
like Rasa \cite{rasa},  the domain of the conversation is organized into dialog states called frames  
 (like state capital) which consists of variables called slots, their values, and prompts to ask the user (for
the values). An example of a slot is the name of a state whose capital the
user wants to find.
There are also pre-defined 
rules using frames and slots to guide DM in the format of Event-Condition-Action declaring what conditions to check for and what corresponding
actions to take. 
There is often a  learning capability to generalize rules in a limited manner (e.g., prompts).
Further, a DM contains several independent modules that are optimized separately, relying on a huge amount of human 
engineering.

\begin{figure}
 \centering
  \includegraphics[width=0.45\textwidth]{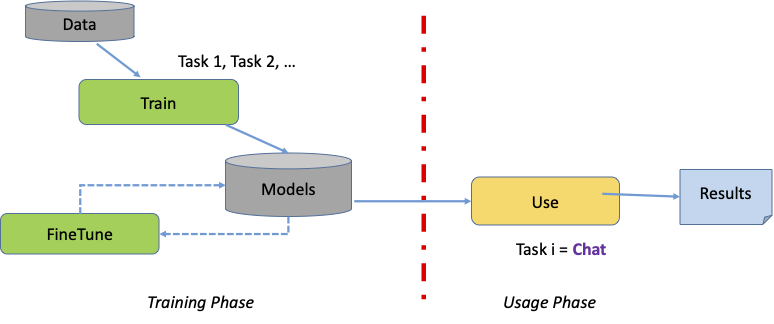}
  \caption{Overview of developing an LLM-based chatbot.}
  \label{fig:chatbot-llm}
\end{figure}

In learning-based approaches, there are many variants. The end-to-end statistical learning approach is to train DM from user utterance to system response without having explicit sub-modules, allowing error signal from the end output of DM to be back-propagated to raw input so that the whole DM can be jointly optimized \cite{e2e-dialog-learning}.
An altogether different approach for DM is using large language models (LLMs) like GPT  \cite{openai2023gpt4}, PaLM \cite{chowdhery2022palm} and LLaMA \cite{touvron2023llama}, which are deep-learning based models trained using large datasets on a set of tasks (see Figure~\ref{fig:chatbot-llm}), and optionally fine-tuned using datasets of a particular domain. An LLM model so trained can be used for a variety of tasks including conversation (chat), i.e., dialog management. ChatGPT and Bard are examples of such chatbots.  When testing such a model, neither the training procedure, consisting of data or tasks, may be known nor the reason why the chatbot generated an utterance - 
thus, truly {\em blackbox} systems. 


Given the long history of chatbots and a plethora of implementation methods, there are surveys \cite{chat-building-survey,dialog-intro,dl-chatbot-survey} and books \cite{chatbot-book,chatbot-book-2} explaining the different approaches for building conversation systems, identifying trade-offs and describing challenges. 
Furthermore, while
current approaches
are effective in building increasingly engaging
chatbots for scenarios with clear goals and in the presence of large training data, the research community recognizes the need to build systems that are collaborative problem solvers \cite{allen-chatbot-complexconv,tyson-aaai18,tyson-ic,cohen-2019-foundations} and can control behavior \cite{dialog-planning-adi,dialog-planning-muise}.
Such systems deal with iteratively refined goals,  need the ability to reason about evolving information and domain and add unique value when the chatbot can take a proactive role in the dialog when it is confident of completing a task with available information.

\section{Evaluation Methods for Chatbots}



Chatbot testing is the process of testing a chatbot to ensure it is functioning correctly and providing an accurate and consistent experience for users. By testing, companies can identify any issues that need to be addressed, such as incorrect responses and errors (functional checks), or slow performance and inappropriate behavior towards users (non-functional checks). Testing also allows to fine-tune chatbot's design and functionality, and assess its overall effectiveness. 

\subsection{Testing Chatbots as Software Systems}


At the least, chatbot testing may be approached as any other  software being tested to meet customer requirements, whether manually or in an automated way \cite{sw-engg-book-Sommerville10,sw-test-ammann2016introduction}.
These methods include unit testing, integration testing, system testing, and acceptance testing. Unit testing is the testing of individual chatbot components, while integration testing is the testing of how those components work together. System testing is a more holistic approach that tests the entire system, including the connections to databases, such as customer, orders, inventory and other. 
Load testing ensures that the chatbot can handle high volumes of traffic without experiencing any issues.
First, the  functional and non-functional requirements need to be converted to system specifications and test plans, and checked for completeness.  
Then, the recommended way to evaluate a system is not at the end, but by checking the system throughout the software development life-cycle. So, {\em unit testing} should be done as each module is getting completed. Subsequently, {\em integration and system testing } is done during  integration followed by  testing  for non-functional requirements before software release. A final {\em acceptance testing} is performed by customers to ensure that all functional and non-functional requirements are met.

The {\em principled} way to build and test software is to come up with a formal specification and evaluate the built system against it. However, it is used in limited, high-stakes situations like chips, aircraft, and military applications. 
There is a long history of modeling the dialog generation process formally, e.g. \cite{young2013pomdp,modeling-conversations-acl}. A simple formulation of a dialog $D$ is as a sequence of utterances [$U^0_h$, $U^0_s$, $U^1_h$, $U^1_s$, ..] between a human ($U^i_h$) and a chatbot system ($U^i_s$), where $i$ is an index into time. Each utterance $U$ consists of word phrases $w_k$. 
However, dialog formulations are rarely used for verifying dialog correctness. If the dialog setting is that of question answering, and each question $Q_k$ and answer $A_k$ pairs are known in a database, a dialog may be considered {\em correct}, if for any ($(U^i_h - Q_k) \preceq \delta$) $\Rightarrow$ ($(U^i_s - A_k) \preceq \delta$), and {\em strictly correct} if $\delta$ = 0. 
That is, for any human utterance containing a question, the system responds with the answer in the database, and the strictly correct system responds with nothing more than the exact answer.
We discuss the potential of formal evaluation later while discussing the ways forward.

\subsection{Testing Chatbots as AI}

As AI systems, chatbots need to be evaluated for the data they use and produce, and the quality of models they use.

\subsubsection{Evaluating  data for privacy and provenance}


As chatbots increasingly become data-reliant, issues 
of what data they are using and how they disseminate insights from it come into question. 
If the data they are using is of poor quality or unreliable, the chatbot will be less effective. 
The first problem with data is the practices used to collect it. With the exception of open data that is published by governments or non-profits for reuse, data collection by industry can be problematic. For example, in the latest report by Mozilla Foundation on data privacy \cite{data-privacy-auto}, 
automobiles scored worst among more than a dozen product categories including fitness trackers, health apps and smart speakers. Among the 25 popular automobile brands in US and Europe the authors reviewed, none   met the minimum privacy standards of Mozilla like control of data and secure storage. This is despite a growing list of regulations on data  like GDPR \cite{chatbot-privacy-law}.
In fact, this was argued as one of the main reasons for lack of adoption of chatbots during Covid-19 \cite{apollo-chatbots,chatbot-privacy-law}.

The second and related problem is of data provenance which refers to the origin and history of data, and how it has been transformed over time. For chatbots, data provenance affects what data they have access to, how they handle data, and how that data is used by downstream applications. 



\subsubsection{Evaluating Models Used }

AI models are  evaluated for technical performance using metrics like
accuracy, recall, and F1 score. When considering custom models, it is important that the most relevant metric is used for evaluation to reveal performance that the user cares about \cite{trust-ml-book}. 
There is also  growing concerns about AI's reliability and trustworthiness with issues like bias (lack of fairness), opaqueness (lack of transparency), 
and brittleness (lack of robust competence)  to perturbations due to omission or commission (adversarial attacks),  regardless of the data they use - be it text \cite{prob-bias-text}, audio \cite{prob-bias-sound} or image \cite{prob-bias-image}. In \cite{dialog-ethics}, the authors explore these issues for chatbots in detail, while \cite{chatbot-recipes} and \cite{chatbot-rating} propose  methods to address them using checklist and trust ratings, respectively.


\subsubsection{Evaluating Large language models (LLMs) }

Another category of models used are
Large Language Models (LLMs) such as the GPT,  PaLM and LLaMA,
which have greatly transformed the field of Natural Language Understanding (NLU) and Generation (NLG).  The use of these LLMs and their chatbot variants like ChatGPT and Bard 
has become a matter of debate due to concerns surrounding ethical implications and their impact on the user's decision-making process. 

The prevalent evaluation strategies can be classified into three major categories based on the aspects they assess: reasoning abilities, fairness and task-specific performance. 
In reasoning, 
LLMs are known to struggle with mathematical reasoning \cite{math-llm-imani2023mathprompter} and causality \cite{kıcıman2023causal}.
In fairness, \cite{nozza-etal-2022-pipelines}  discuss testing methods inspired by software testing where code is continuously pushed into a common repository and tested to ensure the stability of the model. They also discuss the integration of badges, that identify possible social bias issues one might face while using these LLMs, into the development pipeline. 
In the evaluation for task-specific performance, there is a growing list of industry-specific studies (see list at \cite{readings-llm}).
In \cite{lakkaraju2023can}, the authors evaluate LLM-based chatbots, ChatGPT and Bard, on the task of financial advisement by asking 13 questions representing banking products and the interactions between them that included questions in different languages and dialects. The chatbots need to satisfy certain mathematical constraints in order to answer these queries accurately. They concluded that there are critical gaps in providing reliable and accurate responses. In \cite{jalil2023chatgpt}, the authors test whether ChatGPT can answer questions related to software testing accurately. They carefully crafted the queries to test the confidence level and also the explanations provided by ChatGPT along with the accuracy of the response. However, beyond specific results, the learnings from the methodologies used across these different studies are relevant to any new chatbot being built.




\subsection{Testing Chatbots as Human Collaboration Tools}

If humans have to trust a chatbot, they need to be able to understand the latter's output and long-term impact. 
There has been a lot of work in AI explainability \cite{explainable-ai-survey,trust-explainablemetric-survey}. 
 But they have  assumed that
the user understands the AI model’s choices (hence, user-as-developer view) as confirmed
by surveys on usability of explanation\cite{explanability-users-survey}, and not end-users that chatbot users are.


To study long-term impact of any intervention, like technology or medicine, comparative studies like A/B testing or randomized control trial (RCT) are used in technology/ web business and medicine, respectively \cite{ab-rct}. Although there are minor methodological differences, the general idea is to have two groups - a control and a treatment - and to randomly assign experiment subjects to the two groups. 
The two groups are then compared to see if the  impact of the intervention (chatbot here) in the treatment group is more significant than the control group. 

Randomized controlled trials are considered the gold standard for establishing  causal impact of treatment  discounting for unrelated confounders. But their effectiveness depends on the ability to create effective control systems, assign similar subjects to both groups and conducting large scale evaluations in terms of subjects and duration.
In the case of chatbots, very few businesses conduct RCT, with the exception of health where chatbots have undergone small RCTs and shown limited impact \cite{chatbot-health-review}.


\subsubsection{Chatbot Evaluation in Research Community}


There are  ongoing efforts to evaluate chatbots within the research community in competitive settings. Prominent are Dialog System
Technology Challenge (DSTC), a series of competitions whose ninth edition happened in 2021\cite{dstc9-overview}. Each competition has multiple tracks to benchmark chatbots automatically based on various interaction and problem solving capabilities.  Another competition is ConvAI \cite{convai-chatbot-competition} which evaluates conversations based on human evaluation of dialog quality.


\vspace{-0.1in}\section{Testing in Practice and  Open Problems}

We now consider  chatbot testing in mainstream setting of two illustrative settings - digital business and healthcare, and use the context to identify open problems.

\noindent {\bf Chatbots in Business}
Digitalization in business is the process of using digital
technologies to create value for customers. 
Businesses are prominently using chatbots for \textbf{digital sales channel} for prospective opportunities and \textbf{customer service} for retaining existing business.
Digital sales channels lay within the broader concept of digital marketing 
\cite{chaffey2009internet}
and there are many different types of digital sales channels, such as virtual sales agents (chatbots), online marketplaces, and social media platforms.  
Similarly, for customer services, businesses have a variety of digital options including phone, web and chatbots. 

There are a number of stakeholders within the context of digital business using chatbots as shown in Table~\ref{tab:stakeholders}. These include the developers and providers of chatbot technology, the businesses that use chatbots to interact with customers,  the customers, and if relevant, regulators. Each of these groups has a different set of interests and needs, which must be considered when developing and deploying chatbots.
Among these, customers' trust is most important. 
It starts with the technology (chatbot) providing accurate and up-to-date information about the
products or services they are interested in (performance). The tool should be able to work robustly and safely under a range of conditions without jeopardizing user data (reliability). 
The chatbot should also be transparent about the company’s policies and procedures, personable and easy to communicate with, 
and  quick to respond to any questions or concerns that the consumers may have (human interaction).

\begin{table}

  {\small
  
  \begin{tabular} 
{|p{1.5cm}|p{3cm}|p{2.5cm}|}
  \hline
    Stakeholder & Interests &
    Metrics \\
    \hline
     Developers & Build quickly, compatible with internal systems and platforms &
     Dev. time, Bugs, Maint. cost\\ 
     \hline
     
     Businesses & Superior customer experience, cost-effective  &
     NPS, budget \\
      \hline
      
    Customers & Quickly get information, trusting getting the right information  &
    Ease of use, time, trust
    \\
      \hline

        Regulators & Ensure public interest &  complaints, compliance costs 
    \\
  \hline

\end{tabular}
  }
    \caption{Stakeholders of digital business using chatbots}
  \label{tab:stakeholders}
  \vspace{-0.2in}
\end{table}

There are many ideas in literature to make chatbots useful but they do not consider all the stakeholders. For example, 
chatbot treats each user based on their corresponding personality profile \cite{muller2019chatbot}, but it relies on large-scale data collection that users may not want to share and is costly to acquire.
\cite{mozafari2021trust} studied chatbot trust with respect to customer retention. They found that if the chatbot is successful in solving customers’ issues, disclosing the identity of the chatbot will negatively impact customer trust and therefore obstruct retention for highly critical services.

\noindent {\bf Chatbots in HealthCare}
Medical advice chatbots are becoming increasingly popular
as a way to provide quick and easy medical advice. They
are typically used to answer basic questions about medical
conditions and treatments \cite{chatbot-health-review,mental-health-chatbottrust-review}, and can be accessed through a
chat interface on a website or mobile app.
However, critical assessment has shown them to be inadequate 
for emergency assistance like during COVID-19 \cite{apollo-chatbots} or mainstream treatment like breast cancer \cite{breastcancer-health-ai-uk}.

The authors in \cite{seitz2022can} studied diagnostic conversational agents. 
The study identified six trust-building factors in diagnostic chatbots - purpose, reliability, interface design, interaction capabilities, transparency, and relativization.
They found that interpersonal trust approaches do not fully explain the build-up of trust in diagnostic chatbots.
Transparent handling of information and justifications are important to build up trust since trust mostly arises cognitively while underlying the technical character of agents. 
An emerging framework for health chatbots, Chatbot RESET
 \cite{chatbot-reset}, launched in late 2020, provides guidance on how developers, health service providers and regulators can navigate the space.
 It consists of a set of AI and ethics principles as applicable for health use-cases of chatbots and then makes recommendations along the dimensions of optional, suggested and required based on risk to a patient.



\subsection{Open Problems}

From the previous discussions, we now highlight open problems for chatbot testing.

\noindent 1. Lack of 
testing rigor. Given the rich set of evaluation choices, developers and businesses do not always work to maximize customer benefits and focus on their narrow metrics (Table ~\ref{tab:stakeholders}). 
In fact, there is no common testing framework to guarantee any degree of trust in a system - a sharp contrast to testing frameworks in software systems.

\noindent 2.  Lack of test reproducibility. Most chatbots deployed in business settings are dynamic systems whose behavior changes with usage and time. However, to fix a software system, developers need to reproduce reported issues.  

\noindent 3. Not understanding long-term
impact. When businesses introduce technology like a chatbot into production, they may get short-term gains in productivity (e.g., lesser need for human agents) but this intervention may have an adverse long-term impact (e.g., customers going away to competitors due to the perception of poor customer service). If chatbots have to be used in trust-critical applications, their unintended consequences cannot be ignored from evaluation.

\noindent 4. Lack of communication with end-users and stakeholders. 
A chatbot is often delivered to users without any information about its weaknesses or potential issues. For example, how information was used or how uncertain the system is in its output will help a user work collaboratively to the system's strengths without unduly trusting it.

\noindent 5. Inclusive to diverse user groups. As chatbots are deployed to interact with people of different backgrounds and skills, they should be tested for inclusivity.  

\section{A Path Towards  Trustworthy Chatbots}

We now discuss some promising areas for  future research to address open problems and make chatbots trustworthy.

\noindent {\bf Modeling as Dynamical Systems for Evaluation}
An unexplored area is 
modeling chatbots as dynamical systems and using
the relevant systems and control-theoretic tools suitable for their analysis and evaluation. 
Consider one unit time $i$ to consist of an utterance by the human user $U^i_h$ - the \emph{prompt}, and an utterance by the system $U^i_s$ - the \emph{response}. The state of the system at each turn $i$ is denoted by $U^i_s$. 
Often, the system's utterance is in response to a prompt, and hence, we call $U^i_h$ as the exogenous input to the system. Depending on the type of the chatbot design, the state space of the system could be finite or infinite. For example, for a chatbot designed as a rule-based system, the possible responses are constrained to be from a finite set of answers unlike a system designed using LLMs, in which case the state space is infinite. We define the set $\Phi_s$ to include all possible utterances (the states) of the system. The input space, i.e., the set of all possible user prompts, is defined as $\Phi_h$. 

We can then view the chatbot system as a high-dimensional Markov decision process such that the probability of a state transitioning from $U^{i}_s$ to $U^{i+1}_s$ can be represented as 
$P(U^{i+1}_s,U^{i}_s) =  \mathbb{P}(U^{i+1}_s = U|U^{i}_h,U^{i-1}_h,U^{i-2}_h,\ldots, U^{i-m}_h, U^{i}_s,U^{i-1}_s,U^{i-2}_s,\ldots, U^{i-m}_s)$,
where $m$ is the \emph{memory} of the MDP or the context-window. Alternatively, one can model this dynamics as a nonlinear auto-regressive moving average model with exogenous inputs given by
\begin{align}\label{eq:narmax}
   U^{i+1}_s = \mathcal{F}(U^{i}_h,U^{i-1}_h,U^{i-2}_h,\ldots, U^{i-m}_h,\\ U^{i}_s,U^{i-1}_s,U^{i-2}_s,\ldots, U^{i-m}_s)+\mathcal{G}(e^i), \nonumber
\end{align}
where $\mathcal{F}$ is some nonlinear map and $\mathcal{G}(e^i)$ models uncertainties of the system and possible noise at time $i$.

Evaluating the correctness of a chatbot, from a dynamic system perspective is to characterize the convergence property of \eqref{eq:narmax}, i.e., check if $U^{i+1}_s$ reaches a set $\Phi_s^c$ consisting only of \emph{correct} utterances for a given prompt. Here the \textit{measure} of the set $ \Phi_s^c$ is analogous to the $\delta$ defined earlier. In a dialog, the frequency with which the chatbot system stays within the $\Phi_s^c$ can be used to quantify system performance. Ideally, the set $\Phi_s^c$ should be a \textit{forward-invariant set} for the system in a dialog, i.e., the state of the system after reaching this set at turn $i=t$ will remain in this set for any $i>t$. The \textit{reliability} of the chatbot can be characterized based on the stability of the system \eqref{eq:narmax}, i.e., the ability to reach the same state $U \in \Phi_s$ for a given prompt $U_h \in \Phi_h$. A stable dynamical system can be asymptotically stable in the mean or exhibit bounded stability. In the first case, the state may reach $U$, exactly, every time the prompt $U_h$ is presented, while in the second case, it may stay `closer' to $U$ each time, where the closeness of the response with $U$ can be quantified using distance metric in an embedding space or hop distance using knowledge graphs, or human evaluator feedback. For instance, in the example presented in the introduction, the responses to the capital of Alabama can be used to characterize the system as one exhibiting bounded stability. However, such experiments should be repeated several times to perform Monte Carlo analysis, which is essential to characterize the stability of the stochastic system in \eqref{eq:narmax}. Note that the set $ \Phi_s^c$ can be defined in the dataset or the correctness can be quantified by human evaluators, which can be viewed as a noisy reward/reinforcement signal. Similar to $\Phi_s^c$, unsafe utterances for a domain-specific dialog can be grouped together as a set $\Phi_s^{un}$ and the system can be engineered to choose a response $U$ for a given prompt such that $U\in\Phi_s^c \wedge U \notin \Phi_s^{un}$. To fully exploit the dynamic system viewpoint, one may employ various \textit{system identification} techniques to probe the system \eqref{eq:narmax} to characterize the drift-dynamics $\mathcal{F}$ and the uncertainties $\mathcal{G}$ using only the observations (prompts and responses).

This dynamic modeling framework is framed in such a way that the prompts are viewed as \emph{actions}, affecting the evolution of the chatbot states. This hints at the use of classical RL-based planning and learning algorithms for prompt engineering. Secondly, the chatbot engaged in a dialog, especially in the format of a debate or negotiation, can be optimized in a game-theoretic framework, where the user and the system are engaged in a competitive dynamic game. Evaluating chatbots for its safety constraints, bias, and correctness in such competitive setting could yield insightful results - analogous to stress testing a physical system or testing while performing high-stakes task such as playing a game of diplomacy (e.g., Cicero \cite{meta2022human}).


\noindent  {\bf A Trust-focused Chatbot Evaluation Framework}
We need a trust-focused chatbot evaluation framework that is inspired by  lessons of software engineering and pioneering domain-specific attempts like Chatbot RESET\cite{chatbot-reset}. 
To overcome the problem of reproducability of chatbot tests, the framework should
include ability to record a testing session and be able to reproduce observed behavior. In this regard, 
there is prior work that can be adapted in the context of knowledge discovery on how to automate recording and sharing of workflows capturing necessary steps to find a website by interacting with a search  engine CoScripter\cite{coscripter-sharing-workflows}. 
The framework could also guide developers 
on how to choose appropriate fairness metric(s) from the rich choices available \cite{fairness-defs} and make this evaluation an essential part of development process.
Finally, the framework should support design and analysis of user-impact RCT studies to make them routine.  



\noindent {\bf User-focused Communication of Evaluation Results Via Trust Certificates}
The idea of certification assumes that informed users can make the best decisions that impact them.
They work by using the AI property of interest, assessing it in controlled situations (e.g., instances), assigning appropriate certificates, and using it to communicate behavior to users.
In \cite{ai-text-rating-summary,chatbot-rating}, we take a blackbox approach to assess and rate text-based AI.
In \cite{dynamic-certification-autonomous-cacm}, the authors discuss certifying autonomous dynamical systems. In this context, a precise, universal property is not specified. Instead, a list of extensible conditions is given as contexts, each with its corresponding set of properties.



\section{Conclusion}

Building trustworthy chatbots by improving current evaluation procedures is the urgent need of the field.
In this paper, we comprehensively surveyed the problem of evaluating collaborative assistants, for promoting user's trust and confidence. We discussed approaches for building and evaluating chatbots for business settings, identified open problems and explored promising ideas to address them. 

\section{Acknowledgements}

We will like to thank Sara Rae Jones for helping with chatbot evaluation expertiments. We acknowledge funding support from Cisco Research and South Carolina Research Authority (SCRA). 


\bibliography{references/bips_references,references/tarmo_references,references/kausik_references}

\end{document}